\documentclass[aps,prl,twocolumn,showpacs]{revtex4}

\usepackage{amsmath}
\usepackage{amssymb}
\usepackage{graphics}
\usepackage{graphicx}

\begin{document}

\title{Generating particle-like scattering states in wave transport}
  \author{Stefan Rotter}
\thanks{Corresponding author: stefan.rotter@tuwien.ac.at}
\affiliation{Institute for Theoretical Physics, Vienna University of
  Technology, A--1040 Vienna, Austria, EU}
\author{Philipp Ambichl}
\affiliation{Institute for Theoretical Physics, Vienna University of
  Technology, A--1040 Vienna, Austria, EU}
\author{Florian Libisch}
\affiliation{Institute for Theoretical Physics, Vienna University of
  Technology, A--1040 Vienna, Austria, EU}

\date{\today}
 
\begin{abstract}
  We introduce a procedure to generate scattering states which display
  trajectory-like wave function patterns in wave transport through
  complex scatterers. These deterministic scattering states feature
  the dual property of being eigenstates to the Wigner-Smith
  time-delay matrix $Q$ and to the transmission matrix $t^\dagger t$
  with classical (noiseless) transmission eigenvalues close to 0 or
  1. Our procedure to create such beam-like
  states is based solely on the scattering matrix and
  successfully tested numerically for regular, chaotic and disordered
  cavities. These results pave the way for the experimental
  realization of highly collimated wave fronts in transport through
  complex media with possible applications like secure and low-power
  communication.
\end{abstract}

\maketitle

The scattering of waves through complex systems is a central subject
in physics occurring on a variety of length and time scales. Coherent
electron transport through mesoscopic systems, light transmission
through optical devices as well as all matters related to room
acoustics are just a few examples of this kind. Recently, enormous
experimental progress has been made in the ability to determine the
system-specific scattering matrix of such complex systems either
explicitly \cite{Popoff2010Measuring} or implicitly by methods like
adaptive wave-front
shaping \cite{Vellekoop2008Universal} and optical phase conjugation
\cite{Yaqoob2008Optical}. These advances have led to spectacular
results for complex scatterers which could be made
transparent \cite{Vellekoop2008Universal,Yaqoob2008Optical} or put to use for
focusing an incident wave on a spot size below the diffraction
limit \cite{Vellekoop2010Exploiting}.

Common to all such applications is the aim to employ the information
stored in the scattering matrix to create scattering states with
specific properties. A very fundamental property a scattering state can
have is to follow the particle-like bouncing pattern of a classical
trajectory throughout the entire scattering process
\cite{Ehrenfest1927}.  Such ``classical'' scattering states play a key
role for the wave-to-particle crossover, for 
the emergence of geometrical optics out of physical optics and 
for the breakdown of universality in coherent transport
\cite{Silvestrov2003Noiseless,Tworzydlo2003Dynamical,Jacquod2004Breakdown,Jacquod2006Semiclassical,Rotter2007Statistics}.
However interesting these classical states may be, in complex
scattering geometries they turn out to be
as elusive as the proverbial needle in a haystack.

\begin{figure}[!b]
  \centering
  \includegraphics[draft=false,keepaspectratio=true,clip,%
                   width=0.95\columnwidth]{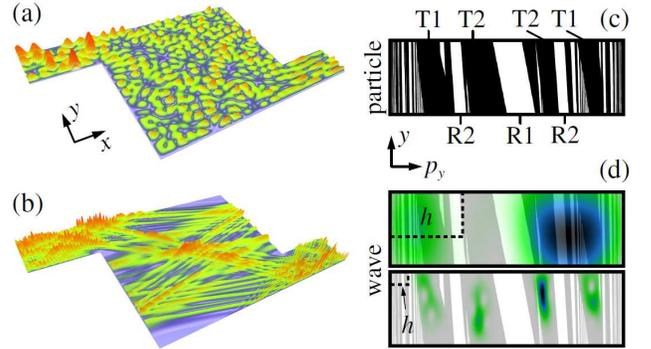}
                   \caption{(Color online) Scattering through a
                     rectangular cavity (flux injected through the
                     left lead of width $d$): (a),(b) Wave function
                     densities of transmission eigenstates
                     $|\tau\rangle$ of $t^\dagger t$ with similar
                     transmission eigenvalues $\tau>0.99$ but
                     different wave numbers: (a) $k\!=\!5.5\pi/d$, (b)
                     $k\!=\!75.5\pi/d$. (c) Classical surface of
                     section, recorded for trajectories which enter at the left
                     lead mouth with vertical position $y$ and
                     transverse momentum $p_y$. The largest of the
                     transmission/reflection bands (black/white) are
                     labeled by T1,T2/R1,R2 (bands which are equivalent
                     in an extended zone scheme are given the same
                     label). (d) Husimi distributions of the states
                     shown in (a) (upper panel) and in (b) (lower
                     panel). The size (area) of the Planck cell $h$ is
                     indicated by dashed black frames and the
                     underlying classical phase space is shown in
                     gray.}
  \label{fig:1}
\end{figure}

In this Letter we propose an operational procedure to generate such
states explicitly. 
Our approach is illustrated with the example
of a two-dimensional rectangular cavity through which waves can be
scattered by two leads attached to the left and right (see
Fig.~\ref{fig:1}). With each lead carrying $N$ open modes the
($2N\times 2N$)--dimensional unitary scattering matrix of this device
has the form,
\begin{equation}
S= \left( \begin{array}{ccc}
r & t'  \\
t & r' \end{array} \right)\,,\label{eq:S}
\end{equation}
where each of the four blocks contains $N\times N$ complex elements
for the energy dependent transmission ($t$) and reflection ($r$)
amplitudes [(unprimed) primed amplitudes designate injection from the
(left) right lead]. The total transmission $T$ through this resonant
cavity is given as $T={\rm Tr} (t^\dagger t)=\sum_{n=1}^N\tau_n$,
where the $\tau_n\in[0,1]$ are the real transmission eigenvalues of
the hermitian matrix $t^\dagger t$. Among the associated eigenstates
$|\tau\rangle$ those with eigenvalues close to $\tau\!=\!0$ or
$\tau\!=\!1$ are termed ``noiseless states'' as they feature a
vanishing contribution to electronic shot noise
\cite{Silvestrov2003Noiseless,Tworzydlo2003Dynamical,Jacquod2004Breakdown,Jacquod2006Semiclassical,Rotter2007Statistics}.
Since all of the desired ``classical'' states with a trajectory-like
bouncing pattern must have such deterministic values of transmission
they are all part of a highly degenerate noiseless subspace associated
with $\tau=0,1$.
Consider in Fig.~\ref{fig:1}a,b two randomly chosen states with
$\tau>0.99$ from this subspace, calculated with the Modular
Recursive Green's Function Method \cite{Rotter2000Modular}.

The first such state (see Fig.~\ref{fig:1}a) was calculated at a low
wave number where only $N=5$ lead modes are open.  To understand the
composition of classical trajectories contributing to this state we
evaluate the Poincar\'e surface of section (PSS) at the entrance lead
junction (see Fig.~\ref{fig:1}c). With transmitted (reflected)
trajectories being shown in black (white) the PSS features a banded
pattern with the individual bands being made up of bundels of
trajectories which all have equivalent bouncing patterns
\cite{Wirtz1997Geometrydependent}. The contributions of different
phase space bands to the state $|\tau\rangle$ in Fig.~\ref{fig:1}a are
revealed by comparing its quantum phase space distribution (Husimi
function) $H(y,p_y)=|\langle\tau|y,p_y\rangle|^2$ with the PSS, where
$|y,p_y\rangle$ is a minimum uncertainty state  at
$y,p_y$. The corresponding plot in Fig.~\ref{fig:1}d (upper panel)
shows that the banded structure of the PSS is not resolved by this
state---in line with the fact that the individual areas of the largest
phase space bands are all smaller than the Planck constant $h$, i.e.,
the lower resolution limit in wave scattering. Rather, this state is
composed of many interfering contributions from both transmitting {\it
  and} reflecting
bands. 

The situation is different when, for smaller wave lengths,
the size of the Planck cell is well below the size of the
largest phase space
bands 
\cite{Silvestrov2003Noiseless,Tworzydlo2003Dynamical,Jacquod2004Breakdown,Jacquod2006Semiclassical,Rotter2007Statistics}.  Consider, e.g.,
the noiseless transmission eigenstate $|\tau\rangle$ with $N=75$ shown
in Fig.~\ref{fig:1}b. We find that the Husimi distribution of this
state [see Fig.~\ref{fig:1}d (lower panel)] is entirely located on
transmission bands, indicating that full transmission is reached here
by resolving the classical phase space.
Since, however, more than one transmission band contribute 
(mostly T1 and T2) clear signatures of classical bouncing patterns are 
still absent in the corresponding scattering wave function (Fig.~\ref{fig:1}b). 
This is a result of the indiscriminate mixing of states in the
degenerate noiseless subspace. This problem may be circumvented by
explicitly constructing scattering states which lie on individual
phase space bands
\cite{Jacquod2006Semiclassical}. In a real
experiment such a protocol, however, meets the problem that the classical
phase space structure is typically unknown for a complex scatterer. A
viable measurement protocol which is based solely on experimentally
accessible quantities like the scattering matrix
\cite{Popoff2010Measuring} would thus be highly desirable.

To resolve the contributions of individual phase space bands we
present an approach based on the observation that all trajectories in
the same band have a characteristic and very similar cavity dwell
time. In analogy to eikonal theory we may thus
``label'' contributions from different bands by their respective dwell
times (or path lengths).  In wave scattering the closest analogues to
classical dwell times are the ``proper delay times'', i.e., the
eigenvalues of the Wigner-Smith time-delay matrix
\cite{Reichl2004Transition},
\begin{equation}
Q=i\hbar\,\frac{\partial
  S^\dagger}{\partial E}S=
i\hbar\left(\begin{array}{ccc}
\dot{r}^\dagger r+\dot{t}^\dagger t & \dot{r}^\dagger t'+\dot{t}^\dagger r'  \\
\dot{t}'^\dagger r+\dot{r}'^\dagger t & \dot{r}'^\dagger r'+\dot{t}'^\dagger t' \end{array}
\right)\,,\label{eq:q}
\end{equation}
where the dots stand for the energy derivative $\partial_E$.  Using
the eigenvalues $q_i$ of $Q$ to lift the unwanted degeneracy in the
noiseless subspace is, however, non-trivial since $Q$ has a different
dimension ($2N\times 2N$) than the transmission matrix $t^\dagger t$
($N\times N$). Accordingly, the eigenstates of $Q$, in general, are
scattering states injected from {\it both} leads, whereas the
eigenstates of $t^\dagger t$ are injected from the {\it left} lead
alone. As shown below, this mismatch is conveniently resolved in the
noiseless subspace where a basis of common eigenstates to both $Q$ and
$t^\dagger t$ can be found.

Due to the hermiticity of $Q$ its eigenstates $|q_i\rangle$ form an
orthogonal and complete set of states, to each of which a real
``proper delay time'' $q_i$ can be assigned.  In the corresponding
matrix representation of this eigenproblem, $Q\,\vec{q}_i^{\,\,\rm
  in}=q_i^{\phantom{.}}\,\vec{q}_i^{\,\,\rm in}$, the $2N$-dimensional
time-delay eigenvectors $\vec{q}_i^{\,\,\rm in}\equiv
(\vec{q}_{i,L}^{\,\,{\rm in}},\vec{q}_{i,R}^{\,\,{\rm in}})$ contain
the complex coefficients of the eigenstates $|q_i\rangle$ in the
flux-normalized basis of {\it incoming} modes in the left
($|n\rangle$) and in the right lead ($|n'\rangle$):
$(\vec{q}_{i,L}^{\,\,{\rm in}})^{\phantom{.}}_n\equiv\langle
n|q_{i}\rangle$ and $(\vec{q}_{i,R}^{\,\,{\rm
    in}})^{\phantom{.}}_{n'}\equiv\langle
n'|q_{i}\rangle$. Correspondingly, the outgoing coefficient vectors
$\vec{q}^{\,\,\rm out}_i\equiv(\vec{q}^{{\,\,\rm
    out}}_{i,L},\vec{q}^{{\,\,\rm out}}_{i,R})$ contain the
coefficients in the basis of outgoing modes: $(\vec{q}_{i,L}^{\,\,{\rm
    out}})^{\phantom{.}}_n\equiv\langle {\mathcal T}n|q_{i}\rangle$
and $(\vec{q}_{i,R}^{\,\,{\rm out}})^{\phantom{.}}_{n'}\equiv\langle
{\mathcal T} n'|q_{i}\rangle$, where ${\mathcal T}$ is the
time-reversal operator of complex conjugation (${\mathcal T}^2\!=\!1$
for spinless scattering). With $\vec{q}^{{\,\,\rm
    out}}_i=S\,\vec{q}^{{\,\,\rm in}}_i$ and $S=S^T$ for systems with
time-reversal symmetry we define an anti-unitarity operator
$\Xi\!=\!{\mathcal T}S\!=\!S^\dagger {\mathcal T}$ which maps the
incoming coefficients of a time-delay eigenstate onto the incoming
coefficients of the corresponding time-reversed state,
$(\vec{q}^{\,\,\rm out}_i)^*\!=\Xi\,\vec{q}^{\,\,\rm in}_i$. As this
operator $\Xi$ commutes with the time-delay operator, $[\Xi,Q]\!=\!0$
(see \cite{epaps}A), any {\it non-degenerate} time-delay eigenstate is
time-reversal invariant (up to a global phase $e^{i\alpha}$,
$\alpha\!\in\!{\mathbb R})$:
$\Xi\,\vec{q}^{\,\,\rm in}_i=e^{i\alpha}\,\vec{q}^{\,\,\rm in}_i$.
For non-degenerate time-delay eigenstates whose incoming flux from one
lead {\it exits} through both of the leads this time-reversal
invariance implies that these states must also have {\it incoming}
flux contributions from both leads. Such $2N$-dimensional time-delay
eigenvectors $\vec{q}_i^{\,\,\rm in}$ can thus not be reduced to an
$N$-dimensional vector with incoming flux from the left lead alone.

This restriction is lifted in the noiseless subspace, where the {\it
  incoming} flux from one lead also {\it exits} through just one of
the leads.  Consider a noiseless time-delay eigenstate with fully
transmitted incoming flux from the left lead, $t^\dagger t
\,\vec{q}_{i,L}^{\,\,{\rm in}}=\vec{q}_{i,L}^{\,\,{\rm in}}$, but no
incoming flux from the right lead, $\vec{q}_{i,R}^{\,\,{\rm
    in}}=\vec{0}$. For this state, $\vec{q}_{i}^{\,\,{\rm
    in}}=[\vec{q}_{i,L}^{\,\,{\rm in}},\vec{0}\,]$, the commutator
$[\Xi,Q]=0$ implies that the time-reversed state,
$\Xi\,\vec{q}_{i}^{\,\,{\rm in}}=[\vec{0},(t\,\vec{q}_{i,L}^{\,\,{\rm
    in}})^*]$, is also a time-delay eigenstate with the same
eigenvalue $q_i$ as $\vec{q}_{i}^{\,\,{\rm in}}$. Being a noiseless
eigenstate of $t'^\dagger t'$ with incoming flux only from the right
lead, $\Xi\,\vec{q}_{i}^{\,\,{\rm in}}$ is clearly orthogonal to
$\vec{q}_{i}^{\,\,{\rm in}}$. We thus find that such ``NOiseless
Time-delay Eigenstates'' (NOTEs) come in pairs of two which together
form the basis of a doubly degenerate subspace associated with the
time-delay eigenvalue $q_i$. We emphasize that, in contrast to the
common eigenbasis of $\Xi$ and $Q$ in this subspace,
$\left[\vec{q}_{i,L}^{\,\,{\rm in}}, \pm (t\,\vec{q}_{i,L}^{\,\,{\rm
      in}})^*\right]/\sqrt{2}$, NOTEs are not time-reversal
invariant. Rather, two NOTEs forming a degenerate pair are the
time-reversed of each other like a classical trajectory and its
time-reversed partner. Focussing now only on NOTEs injected from the
left lead, we can determine their expansion coefficients
$\vec{q}_{i,L}^{\,\,{\rm in}}$ with $Q$ from Eq.~(\ref{eq:q}),
\begin{equation}
  \left( \begin{array}{cc}
      Q_{11}\phantom{\Big|}\! & Q_{12}  \\
      Q_{21}\phantom{\Big |}\! & Q_{22} \end{array}
  \right)\left( \begin{array}{c}
      \!\!\phantom{\Big |}\vec{q}_{i,L}^{\,\,{\rm in}}\phantom{\Big |}\!\!
      \\\!\!\phantom{\Big|}\vec{0}\phantom{\Big|}\!\!
    \end{array} \right)=
  \left( \begin{array}{c}
      \!\!\phantom{\Big|} Q^{\phantom{i}}_{11}\,\vec{q}_{i,L}^{\,\,{\rm in}}\phantom{\Big|}\!\!
      \\\!\!\phantom{\Big|}Q^{\phantom{i}}_{21}\,\vec{q}_{i,L}^{\,\,{\rm in}}\phantom{\Big|}\!\!
\end{array} \right)=
  q_i \left( \begin{array}{c}
      \!\!\phantom{\Big |}\vec{q}_{i,L}^{\,\,{\rm in}}\phantom{\Big |}\!\!
      \\\!\!\phantom{\Big|}\vec{0}\phantom{\Big|}\!\!
\end{array} \right)
\,.\label{eq:q2}
\end{equation}
For the last equality to hold, the following two conditions need to be
fulfilled: (i) $Q^{\phantom{i}}_{11}\,\vec{q}_{i,L}^{\,\,{\rm
    in}}=q^{\phantom{i}}_i\,\vec{q}_{i,L}^{\,\,{\rm in}}$ and (ii)
$Q^{\phantom{i}}_{21}\,\vec{q}_{i,L}^{\,\,{\rm in}}=\vec{0}$.  This
central result of our Letter implies the following operational
procedure to determine the expansion coefficients of NOTEs: In a first
step (i) the eigenstates of the hermitian matrix $Q_{11}$ of dimension
$N\times N$ are calculated. Out of this orthogonal and complete set of
vectors the subset which, according to (ii), lies in the null-space
(kernel) of $Q_{21}$, constitutes the desired set of common
eigenstates of $Q$ and $t^\dagger t$. 
In practice, condition (ii) can be conveniently verified by a
null-space norm $\chi_i=\|Q_{21}\,\vec{q}_{i,L}^{\,\,{\rm in}}\|$
which determines the degree to which the normalized vector
$\vec{q}_{i,L}^{\,\,{\rm in}}$ lies in the null-space of $Q_{21}$. The
quality of a NOTE should be the better the closer this measure
$\chi\in[0,\infty]$ is to zero. As the limiting value $\chi\to 0$ is
only reached for exact NOTEs with wavelength $\lambda\to 0$, we need
to test whether our approach works also for the realistic situation where
$\lambda$ has a finite value. 

\begin{figure}[!t]
  \centering
 \includegraphics[draft=false,keepaspectratio=true,clip,%
                   width=0.95\columnwidth]{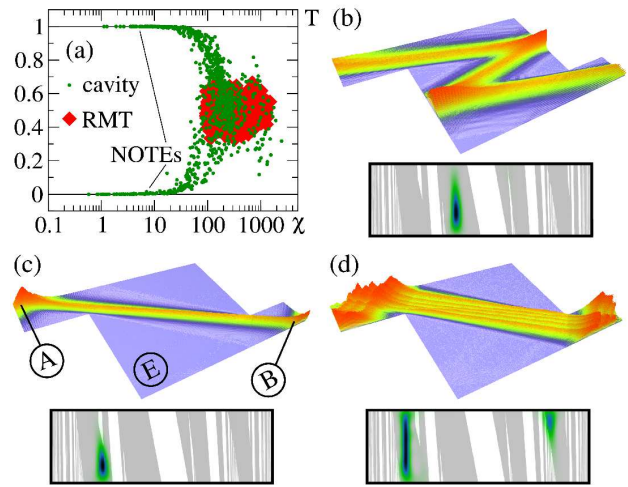}
                   \caption{(Color online) (a) Transmission $T$
                     vs.~null-space norm $\chi$ for eigenstates of the
                     matrix $Q_{11}$ in a rectangular cavity with
                     different lead orientations (green dots). NOTEs
                     with $\chi\to 0$ are noiseless and strongly
                     deviate from RMT (red diamonds, see \cite{epaps}B
                     for details). (b)--(d) Wave function
                     densities for NOTEs calculated with the same
                     scattering matrix data as used for
                     Fig.~\ref{fig:1}b. As demonstrated by the Husimi
                     plots in the bottom panels, each state is located
                     on a single classical phase space
                     band. Null-space projections $\chi$ are (b) $4.7$,
                     (c) $6.3$, (d) $6.9$.  The insets in (c) illustrate
                     the possibility to use NOTEs for transferring
                     information between a sender (A) and a receiver
                     (B) which by-passes a potential eavesdropper
                     (E).}
  \label{fig:2}
\end{figure}
Consider, as a starting point for such a test,
our previous argument that NOTEs can only exist
in the noiseless subspace with $\tau=0,1$. We emphasize that this
requirement, which can similarly not be fulfilled exactly for any finite
value of $\lambda$, does not explicitly enter conditions (i),(ii) from
above. A good indicator for the validity of our approach  
is thus the degree to which NOTEs with finite values of $\lambda$ 
are, indeed, noiseless. For this purpose we calculate 
the eigenvectors $\vec{q}_{i,L}^{\,\,{\rm in}}$ of $Q_{11}$ [see condition (i)] and
verify how the transmission of these states correlates with the
corresponding null-space norm $\chi_i$ [see condition (ii)]. 
We find 
that all eigenstates of $Q_{11}$ which closely fulfill the NOTEs
condition (ii) of low $\chi$-values are, indeed, either almost fully
transmitted, $\tau\approx 1$, or fully reflected, $\tau\approx
0$. This behavior is entirely absent in Random Matrix Theory (RMT)
where no phase space bands exist (a comparison of the data for the
rectangular cavity with RMT is shown in Fig.~\ref{fig:2}a and in 
\cite{epaps}B for other scattering geometries). 
After these consistency checks we test whether NOTEs with very low
null-space norms ($\chi\lesssim 30$, see Fig.~\ref{fig:2}a) display in
their wave functions the anticipated pronounced enhancements around
individual bundles of classical trajectories. Our results based on the
same scattering matrix data as for Fig.~\ref{fig:1}b confirm that
states with such low null-space norms $\chi$ all feature highly
collimated beam-like wave functions (see Fig.~\ref{fig:2}b-d and
\cite{epaps}E). Quite different from arbitrary noiseless states
(Fig.~\ref{fig:1}a,b), NOTEs feature Husimi distributions that do not
mix contributions from different phase space bands, thereby
corroborating the successful operation of our procedure. Without
exception we find that in cases where NOTEs seem to feature
contributions from more than one band (as in Fig.~\ref{fig:2}d) all
these bands belong to a single connected band in an extended zone
scheme (like T1/T2/R2 in Fig.~\ref{fig:1}c)
\cite{Wirtz1997Geometrydependent}.

Our numerical results indicate furthermore that proper delay times
of NOTEs do not only lift the degeneracy of noiseless states located
on {\it different} bands, but that also the small dwell-time
differences between trajectories of the {\it same} band do get
increasingly well resolved in the limit $\lambda\to 0$.
Correspondingly, we find that the proper delay times $q_i$ of NOTEs on
the {\it same} band are characteristically different from each other
(rather than degenerate). NOTEs thus fill individual phase space bands
in a well-controlled fashion. Consider, e.g., the band T1: starting
from the state in Fig.~\ref{fig:2}c the proper delay times and the
transverse quantization of states on this band increase (see, e.g.,
Fig.~\ref{fig:2}d) until, when the band is filled, the null space norm
of states increases substantially (see \cite{epaps}E),
indicating a substantial overlap with phase space outside of the
band. Such an increase in $\chi$-values is often found to be
accompanied by signatures of diffractive scattering at the
sharp lead mouths (see \cite{epaps}E).

\begin{figure}[!t]
  \centering
 \includegraphics[draft=false,keepaspectratio=true,clip,%
                   width=0.95\columnwidth]{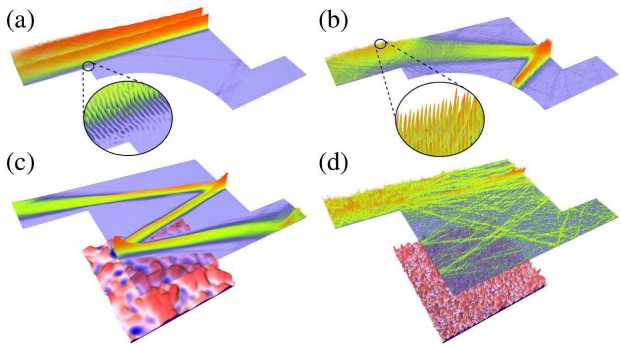}
                   \caption{(Color online) NOTEs at $k=75.5\pi/d$ in
                     cavities (a),(b) with a Sinai-type boundary shape
                     and (c),(d) with a bulk disorder of amplitude
                     $V_0/E=0.1$ (see
                     illustrations). The correlation of the disorder
                     is long-range for (c) with correlation length
                     $kr_c\!=\!30\pi$ and short-range for (d) where
                     $kr_c\!=\!5\pi$. Null-space norms $\chi$
                     are (a) $1.7$, (b) $24.8$, (c) $4.9$, (d)
                     $65.7$.}
  \label{fig:3}
\end{figure}

Since the operational procedure presented here does not rely on any
specific assumptions concerning the type of scattering in a given
system, we also applied it to more complex scattering
geometries. Consider first the Sinai-type billiard structure in
Fig.~\ref{fig:3}a,b which features chaotic classical dynamics due to
scattering at the circular part of the hard wall potential. We find
that NOTEs which do not have any overlap with the curved part of the
boundary (see Fig.~\ref{fig:3}a) have wave functions and proper delay
times as in the rectangular cavity. For comparison, we also show in
Fig.~\ref{fig:3}b a state which bounces off the circular boundary.  As
reflected in its increased null-space norm ($\chi=24.8$) the high
instability of this state's bouncing pattern makes it much harder to
resolve the (small) area of its phase space band. Consider next 
rectangular cavities containing a static disorder potential with
correlation length $r_c$ and an average amplitude $V_0$. By studying
such disordered cavities we find that our approach is restricted to
the limit of weak and long-range disorder, $V_0/E\ll 1$ and $k r_c\gg
1$ (in agreement with the validity criterion for the WKB/eikonal
approximation \cite{glauber} and previous work
\cite{Rotter2007Statistics}). 
Correspondingly, for the same fixed disorder amplitude ($V_0/E=0.1$)
the $z$-shaped scattering state from Fig.~\ref{fig:2}b survives in the
presence of long-range disorder (see Fig.~\ref{fig:3}c), whereas no
such state exists for short-range correlations (see
Fig.~\ref{fig:3}d for the state with the closest time-delay
value and \cite{epaps}B,E for more details).

We have performed additional tests (see \cite{epaps}C) to verify that
all NOTEs which we find within the above
limits 
are associated with individual classical trajectory bundles.
This feature allows for a WKB/eikonal-type ansatz for the transmission
matrix of NOTEs,
$t\approx\vec{q}_{i,R}^{\,\,{\rm out}}\, e^{i{\mathcal
    S}_b(E)/\hbar}\,\vec{q}_{i,L}^{\,\,{\rm in}\,\dagger}$, in which
the only part with a significant energy-dependence is the action
phase, ${\mathcal S}_b(E)=\int_b \vec{k}\,d\vec{l}$, accumulated along
bundle $b$. We show in \cite{epaps}D that this ansatz fulfills the
defining conditions for NOTEs (i),(ii) from above.

We believe that our results open up many interesting possibilities
for the experiment, where the cavities considered here could, e.g., be
an acoustic resonator (like a room) or an electromagnetic scatterer
(like closely spaced buildings). In both these cases the collimated
wave functions associated with NOTEs might allow to transfer
information between a sender (A) and a receiver (B) such that the
power to generate the signal is minimized and the transmitted signal
is kept out of reach of an eavesdropper (E) (see illustration in
Fig.~\ref{fig:2}c). In this sense NOTEs offer clear advantages over
arbitrary noiseless scattering states that do not display such beam-like
wave functions in general (see, e.g., Fig.~\ref{fig:1}b).
NOTEs may also have interesting connections to phenomena in 
closed/decaying systems \cite{vergini2010,kopp2010,hakan2002}.

In summary, we present an operational procedure for constructing
scattering states which follow classical bouncing patterns in coherent
transport through cavities or complex scattering landscapes. In
analogy to WKB/eikonal theory we find that such
ray-optical/beam-like scattering states are determined by the
condition of a fixed scattering time-delay. Our procedure
is generally applicable to different types of
wave scattering (acoustic, electro-magnetic, quantum etc.) and relies
solely on the knowledge of the scattering
matrix. 

\begin{acknowledgments}
  We wish to thank F.~Aigner, J.~Burgd\"orfer, A.~Cresti, and
  A.~Foerster for helpful discussions. Support by the WWTF and computational
  resources by the Vienna Scientific Cluster (VSC) are gratefully acknowledged.
\end{acknowledgments}

\end{document}


\title{Supplemental material for\\
"Generating particle-like scattering states in wave transport"}
  \author{Stefan Rotter, Philipp Ambichl, and Florian Libisch}
\affiliation{Institute for Theoretical Physics, Vienna University of
  Technology, A--1040 Vienna, Austria, EU}

\date{\today}
 

\maketitle
 \renewcommand{\theequation}{E.\arabic{equation}}
	
\section{A.~Time-reversal operator}
In the main article we introduce the anti-unitary time-reversal operator
$\Xi={\mathcal T}S$ which maps the incoming coefficients of a 
scattering state onto the incoming coefficients of the corresponding 
time-reversed state. For time-reversal invariant systems like the ones
we study the scattering matrix $S$ is unitary symmetric such that
the time-reversal operator $\Xi$ can also be written as follows:  
$\Xi=S^\dagger{\mathcal T}$. This identity can be conveniently
used to show that $\Xi$ commutes with the Wigner-Smith time-delay 
operator $Q$ which, due to its Hermiticity, satisfies the  
following identities: $Q=i\hbar(\partial_E S^\dagger) S=-i\hbar
S^\dagger (\partial_E S)$. In particular, we have:
\begin{eqnarray}
[Q,\Xi]&=&\left\{i\hbar(\partial_E S^\dagger) S\right\}S^\dagger {\mathcal T}-
S^\dagger {\mathcal T}\left\{-i\hbar S^\dagger (\partial_E S)\right\}=\nonumber\\
&=&i\hbar\left\{(\partial_E S^\dagger) S S^\dagger {\mathcal T}-S^\dagger 
S{\mathcal T}(\partial_E S)\right\}=\nonumber\\
&=&i\hbar\left\{(\partial_E S^\dagger) {\mathcal T}-(\partial_E S^\dagger) {\mathcal T}\right\}=0\,.
\end{eqnarray}

\section{B.~Statistical Analysis of NOTEs}
Since we expect the presence of NOTEs in a given scattering system to 
depend on non-universal, system specific scattering processes, we
should find no NOTEs in a system described by an entirely random
scattering matrix as in Random Matrix Theory (RMT). Whereas RMT is
thus not a suitable framework for studying NOTEs, the
differences that appear between an RMT description and a full
numerical solution of a non-universal scattering system will be
very instructive for characterizing NOTEs in detail.
\subsection{Random Matrix Model}
For a suitable RMT model of scattering which provides also the
Wigner-Smith time delay matrix we follow the so-called ``Heidelberg
approach'' reviewed, e.g., in \cite{fyodorov}. This approach links the
scattering matrix $S$ of a system with the Hamiltonian $H$ of
the corresponding closed cavity through the coupling matrix $W$ between
the discrete bound states in the cavity and the continuum in the
attached leads,
\begin{equation}\label{eq:1}
S(E)=I-2\pi i W^\dagger\frac{1}{E-H+i\pi W W^\dagger} W\,.
\end{equation}
For time-reversal invariant systems the Hamiltonian matrix $H$ is
taken from the Gaussian orthogonal ensemble ($\beta=1$). The mean
spacing of the energy eigenvalues of $H$ is adjusted to match the
corresponding mean level spacing $\Delta E=2\pi/A$ of our cavities
with size $A$. (Note that the chaotic cavities we study have a smaller
scattering area $A$ and are thus compared with a different RMT
ensemble than the regular and the disordered cavities.) 
The mean absolute value of the random coupling matrix
elements of $W$ is adjusted such as to produce a scattering matrix $S$
with, on average, balanced total transmission and reflection, $\langle
T\rangle\approx \langle R\rangle\approx \langle T'\rangle\approx
\langle R'\rangle\approx N/2$ \cite{pichugin}.

For calculating the Wigner-Smith time-delay matrix based on
Eq.~(\ref{eq:1}), $Q=i\hbar(\partial_E S^\dagger) S$, we make the
usual assumption \cite{fyodorov} that the coupling matrix elements
in $W$ are only weakly energy-dependent such that the derivative
$\partial_E S^\dagger$ affects only the explicit energy dependence in
the denominator of Eq.~(\ref{eq:1}). To check our results we have
verified that the Wigner-Smith time-delay matrix $Q$ obtained in this
way gives rise to eigenvalues $q_i$ which are statistically
distributed according to the RMT predictions put forward in
\cite{sommers} (not shown).

\begin{figure}[!t]
\includegraphics[draft=false,keepaspectratio=true,clip,%
                   width=
                   \columnwidth]{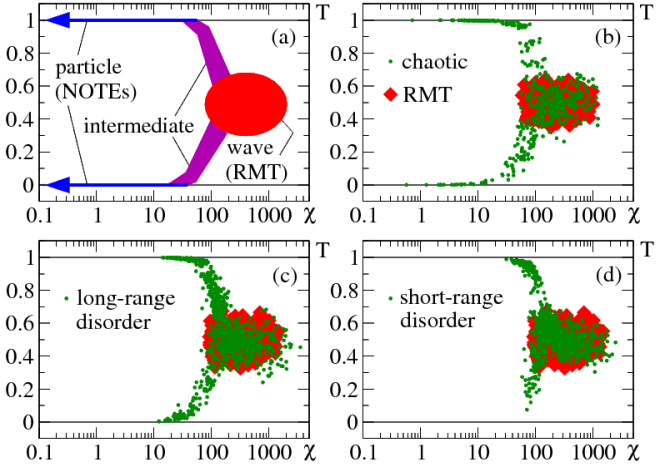}
                   \caption{Transmission versus null-space norm $\chi$
                     for eigenstates of the submatrix $Q_{11}$ of the
                     Wigner-Smith time-delay matrix $Q$ (wave number
                     $k=45.5\pi/d$). (a) Schematic of the results
                     shown in Fig.~2a (main text) to illustrate the
                     crossover between particle/classical and
                     wave/quantum scattering. (b)-(d) Same as Fig.~2a
                     (main text) for (b) the Sinai-type cavity with
                     chaotic classical dynamics as well as for weakly
                     disordered cavities ($V_0=0.1 E$) with (c)
                     long-range ($kr_c=15\pi$) and (d) short-range
                     ($kr_c=2.5\pi$) disorder. The red diamonds show
                     the data obtained from the RMT calculations from
                     altogether 20 randomly assembled $S$-matrices
                     (for better visibility we use an increased symbol
                     size for the RMT data).  Strong deviations
                     between the cavity and the RMT data are observed
                     which point to the presence of NOTEs with small
                     null-space norms $\chi\to 0$ and noiseless
                     transmission $T(1-T)\to 0$. 
                     The presence of a disorder potential (in
                     particular, with short-range correlation) shifts
                     the cavity data towards the RMT distribution and
                     strongly reduces the amount and the quality of
                     NOTEs.  }
\label{fig:1}
\end{figure}

\subsection{Comparison of numerical data with RMT}
It is now of interest to compare the results obtained with this RMT
approach with those from our numerical calculations for the regular,
chaotic, and disordered cavities. The RMT data was produced based on
an ensemble with altogether 20 random scattering matrices following
Eq.~\eqref{eq:1}. To obtain a comparatively large ensemble of
statistical data also for the cavities, we performed the numerical
wave calculations with
20 different disorder realizations (for the disordered cavities) and
with 20 different lead configuations (for the regular and the chaotic
cavity). In the latter case, we shifted the right lead in equidistant
steps from the bottom to the top position (the left lead was kept
fixed at the top position). To counterbalance the increased numerical
effort necessary for calculating such ensembles of cavities, we
reduced the wavenumber from $k=75.5\pi/d$ (as used for the plots in
the main text) to $k=45.5\pi/d$.

In both the RMT and the cavity calculations we focused, in particular,
on the eigenstates $\vec{q}_{i,L}^{\,\,{\rm in}}$ of $Q_{11}$ which,
according to condition (i) in our procedure (see main text), are the
key ingredients for NOTEs. In Fig.~E.1 we plot the total
transmission of these states as a function of the corresponding
null-space norm $\chi=\|Q_{21}\vec{q}_{i,L}^{\,\,{\rm in}}\|$ which,
according to condition (ii) measures whether a state can be identified
as a NOTE. In this plot already a number of important features become
apparent: the data from the RMT calculations (red diamonds) accumulate
in a very restricted region of the parameter space: the norm $\chi$ of
the RMT states stays within a restricted interval of $\chi$ and their
transmission clusters very strongly around the mean transmission
$T=0.5$. This situation is quite different for the cavities: whereas
the upper bound of the norm $\chi$ is comparable to the corresponding
RMT value, the lower bound of $\chi$ is more than two orders of
magnitude smaller than in RMT, suggesting a strongly non-universal
behavior. Furthermore we observe that states with very low values of
the norm $\chi$ are all clustered around the noiseless transmission
values very close to $T=0$ and $T=1$. 

This observation demonstrates
that, indeed, eigenstates of $Q_{11}$ with a vanishing nullspace norm
[$\chi\to 0$] have noiseless transmission [$T(1-T)\to
0$]. Equivalently, each scattering state injected from just the left
lead [i.e., with $\vec{q}_{i,R}^{\,\,{\rm in}}=0$] which is a
time-delay eigenstate is found to be either fully transmitted or fully
reflected. NOTEs are thus, indeed, noiseless. We can further write the
transmission $T$ of a NOTE as $T=\sum_i |c_i|^2\tau_i$, where the
$\tau_i$ are the transmission eigenvalues of $t^\dagger t$ and the
$c_i$ are the expansion coefficients of the NOTE in the eigenbasis of
$t^\dagger t$. From this we conclude that the noiseless transmission
values of NOTEs, $T\approx 0$ ($T\approx 1$), imply that NOTEs can
only have non-negligible expansion coefficients $c_i$ for those
transmission eigenvalues $\tau_i$ which are themselves noiseless,
$\tau_i\approx 0$ ($\tau_i\approx 1$). NOTEs can thus only be composed
of noiseless transmission eigenchannels. 

We emphasize that the strong correlation between time-delay
eigenstates and noiseless scattering (which does not explicitly enter
our construction of NOTEs) persists here in the limit of finite
channel numbers where neither the null-space norm $\chi$ is exactly 0
nor the transmission $T$ is exactly noiseless for any state.
To highlight this strong correlation we show in Fig.~E.2 the
null-space norm $\chi$ of $Q_{11}$ eigenstates as a function of $\sigma=4
T(1-T)$, which measure for noiseless scattering can vary between 0 and
1. For all states which display non-universal behavior (deviating from
the RMT-result) a strong positive correlation between the two measures
$\chi$ and $\sigma$ is observed, suggesting that we can use these two
measures interchangeably for testing the quality of a NOTE. We have
checked explicitly that the statistical spread in the correlation
between $\chi$ and $\sigma$ (see Fig.~E.2) is further reduced when
considering states on the same phase space band.

\begin{figure}[!t]
\includegraphics[draft=false,keepaspectratio=true,clip,%
                   width=
                   \columnwidth]{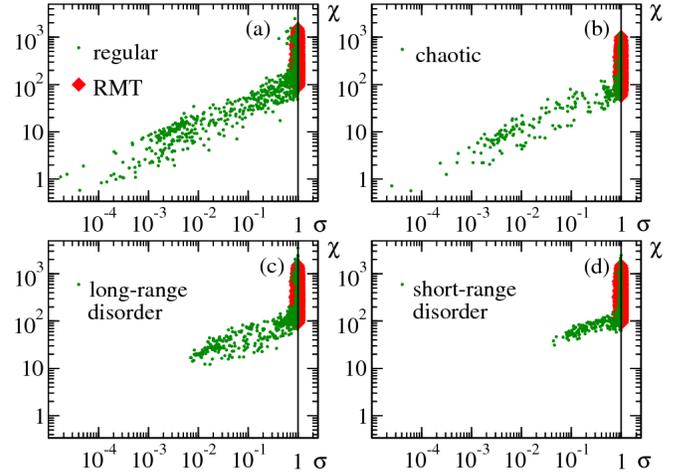}

                   \caption{Null-space norm $\chi=\|Q_{21}\vec{q}_{i,L}^{\,\,{\rm in}}\|$ versus 
                   noiseless norm $\sigma=4T(1-T)$ for the four different cavity types investigated.
                   The same data set and color coding is used as in  
                   Fig.~E.1. A strong positive correlation between the two norms 
                   $\chi$ and $\sigma$ is visible for all data points that are strongly
                   non-universal, i.e., for parameter values that are different from  
                   those where the RMT results cluster (see red diamonds).}
\label{fig:2}
\end{figure}

The statistical data shown in Fig.~E.1 and Fig.~E.2 also demonstrate
that NOTEs are completely absent in the RMT description along
Eq.~\eqref{eq:1}. This observation is perfectly in line with the
expectation that the presence of NOTEs requires extended bands in
phase space. As such non-universal features are clearly absent in the
RMT description, this model does not give rise to any NOTEs. Rather,
the eigenstates of the random matrix $Q_{11}$ do not constitute a
preferred basis in terms of transmission (as opposed to the
eigenstates of $t^\dagger t$). Their transmission values thus cluster
around the average value $T\approx 0.5$ and are strongly suppressed
near the noiseless values $T\approx 0,1$. As discussed in the main
part of the paper, one can show by employing the time-reversal
operator $\Xi$ that such states which lie outside the noiseless
subspace can never be time-delay eigenstates when injected from just
one of the leads.  Correspondingly, the $Q_{21}$ null space norm
$\chi$ of these RMT states always remains larger than the (small)
$\chi$-values of NOTEs.

\begin{figure}[!t]
\includegraphics[draft=false,keepaspectratio=true,clip,%
                   width=
                   \columnwidth]{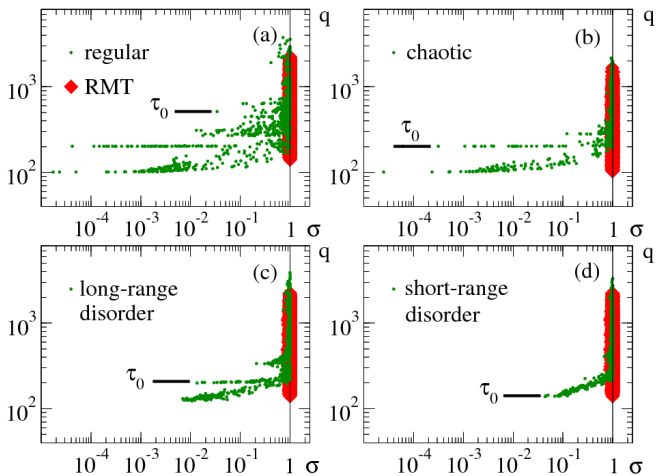}
                   \caption{Eigenvalues $q_i$ of $Q_{11}$
                     versus noiseless norm $\sigma$ of the
                     corresponding eigenvectors
                     $\vec{q}_{i,L}^{\,\,{\rm in}}$. The same data set
                     and color coding is used as in Fig.~E.1. As
                     expected from our model, all states with a small
                     noiseless norm $\sigma$ correspond to short
                     delay-time values $q_i$. A clustering of these
                     $q_i$-values around the classical values
                     belonging to individual phase space bands is
                     clearly observed. Note, e.g., that a delay time
                     of 100 (200) corresponds to a single (double)
                     traversal of the cavity along the upper cavity
                     boundary in those cavity configurations where the
                     right lead is (is not) located at the top-most
                     position.  The maximum delay-time $\tau_0$ where
                     such non-universal deviations with $\sigma<0.1$
                     occur are indicated by the black horizontal
                     bars. The values of $\tau_0$ are reduced in chaotic
                     vs.~regular ballistic cavities and in short-range
                     vs.~long-range disordered cavities.}
\label{fig:3}
\end{figure}

From the above results we conclude that the noiseless eigenstates of
$Q_{11}$ are NOTEs for which the corresponding eigenvalues $q_i$ of
$Q_{11}$ are time-delay values. Following our model we expect that
these time-delay eigenvalues correspond to the time spent in the
cavity when scattering {\it along a given trajectory bundle} (see
section C for more details).  Consequently, all states that belong to
the same bundle should feature similar values of $q_i$, which we
expect to lead to a clustering of $q_i$-values around the classical
time delay values of those trajectory bundles which can be fully
resolved by a scattering state. As, in turn, the degree of this
resolution is measured by the noiseless norm $\sigma$ (or,
alternatively, by the null-space norm $\chi$) it is instructive to
plot the eigenvalues $q_i$ of $Q_{11}$ as a function of these norms
$\sigma$ (or $\chi$).  The resulting plot shown for the noiseless norm
$\sigma$ in Fig.~E.3 nicely confirms the expected behavior: we first
observe that no states with both very long time-delays {\it and} small
$\sigma$-values are found, in correspondence with the expectation that
NOTEs can only form on short-lived/stable phase space bands. States
with small norm values $\sigma$ show the most pronounced deviations
from RMT and correspond exclusively to short-lived states with
time-delay values on the lower end of the entire distribution.  For a
single cavity these non-universal $q_i$-values cluster along
horizontally elongated regions, corresponding to states on the same
phase space band. For such states the time-delay values $q_i$ are
similar, but the norms $\sigma$ vary considerably, depending on the
overlap of a state with regions outside the band. In Fig.~E.3 such a
clustering of values is still visible but blurred due to the different
lead configurations that enter the statistical data set. Those NOTEs
out of this set with the lowest values of $\chi$ correspond to
scattering states which propagate solely along the upper cavity
boundary where the focussing is enhanced by the top cavity wall. In
contrast, all of these non-universal features are drastically reduced
for states with a norm $\sigma$ near the maximum value 1.

\subsection{Signatures of regular/chaotic/disorder scattering}

Since NOTEs are highly non-universal and system-specific scattering
states, their number and their quality should depend significantly on
the type of scattering present in a given scattering region. To
investigate this issue in more detail we calculated all of the above
plots for the regular, chaotic and disordered cavities,
respectively. As a measure for the influence of these different
scattering mechanisms we introduce a critical time scale $\tau_0$
which represents the maximal time up to which 
time-delay eigenvalues of a given quality ($\sigma_0=0.1$) exist for a
given system. As this time scale $\tau_0$ measures for how long
scattering states follow the classical bouncing patterns of NOTEs (see
section C) $\tau_0$ can loosely be associated with the Ehrenfest time (see
Refs.~[5-10] in the main paper). 

In all of the above plots we consistently find that the strongest
deviations from RMT occur for the regular cavity which,
correspondingly, also features the largest value of $\tau_0$ (see
black horizontal bars in Fig.~E.3). Note that for the regular cavity
deviations from RMT do persist to a certain degree also for states
with time-delays longer than $\tau_0$, indicating that in regular
cavities also very long-lived states may carry non-universal
features. Comparing these results with those for the chaotic cavity we
observe the following interesting behavior: whereas states with the
shortest time-delays feature very similar values of the norm $\sigma$
in the regular and the chaotic cavities, $\tau_0$ is considerably
shorter for the chaotic scatterer than for the regular one. This can
be intuitively explained by the fact that short-lived NOTEs which do
not have any overlap with the circular part of the boundary in the
chaotic dot are the same as the corresponding states in the regular
cavity. Those very stable states corresponding to the smallest
$\sigma$ values thus appear in both cavity types with similar
parameter values. Longer-lived states do eventually also explore the
circular boundary part in the chaotic cavity which leads to increased
wave spreading and thus to a reduced value of $\tau_0$.  In the cavity
with long-range disorder the situation is again 
conspicuously different from
the chaotic cavity, although both types of cavities give rise to
classical chaotic scattering: in the long-range disordered cavity even
the most short-lived states are drastically affected by the disorder
since it is contained in the bulk and affects all scattering states
(in contrast to the chaotic cavity where the circular boundary part
can be avoided). Correspondingly, the minimum value of the norm
$\sigma$ is much higher for the disordered cavity than for both the
regular and the chaotic cavity. When changing the correlation length
of the disorder to smaller length scales (with the disorder amplitude
remaining unchanged at $V_0/E=0.1$) the effect of the disorder is
further increased: here even the states with the shortest delay time
have $\sigma$ values which are three orders of magnitude larger than
in the ballistic cavities without disorder. Also $\tau_0$ is here much
smaller than in all other cavity types. This behavior can be
attributed to the effect of stochastic scattering which is induced
whenever the correlation length of the disorder is of comparable
magnitude or smaller than the wavelength.

\subsection{NOTEs and general time-delay eigenstates}
In the main part of the article we argue that NOTEs should appear as a
special subset of eigenstates of the Wigner-Smith time-delay operator
$Q$. 
A convenient way to test this relation between NOTEs and general
time-delay eigenstates is to compare the eigenvalues and eigenvectors
of the submatrix $Q_{11}$ (which we use to generate NOTEs) with those
of the general Wigner-Smith time-delay matrix $Q$. In particular, we
expect that the eigenvectors and eigenvalues belonging to a NOTE
(i.e., for which $\chi,\sigma\to 0$) should emerge for both matrices,
whereas the remaining eigenstates (with higher values of $\chi$,
$\sigma$) may again be close to the corresponding RMT values for each
of these two matrices. To check whether this expected behavior is,
indeed, fulfilled consider Fig.~E.4: this plot shows
the eigenvalues $q_i$ as a function of the noiseless norm $\sigma$
obtained here from the entire matrix $Q$ rather than only from
$Q_{11}$ as in Fig.~E.3. The two sets of eigenstates from $Q$ and $Q_{11}$
are of different size since the eigenvectors $\vec{q}_{i}^{\,\,{\rm
    in}}$ of $Q$ contain the incoming mode coefficients from both
leads, $\vec{q}_{i}^{\,\,{\rm in}}=(\vec{q}_{i,L}^{\,\,{\rm
    in}},\vec{q}_{i,R}^{\,\,{\rm in}})$, not just from the left lead
alone. We thus calculate the norm $\sigma=4T(1-T)$ for Fig.~E.4 just
based on the transmission $T$ associated with the (normalized)
injection coefficients from the left lead, i.e., with
$\vec{q}_{i,L}^{\,\,{\rm in}}/\|\vec{q}_{i,L}^{\,\,{\rm in}}\|$.
Since we expect NOTEs to appear in the eigenvectors of $Q$ as a near
degenerate doublet (see main text) the normalization removes here the
effect of mixing between these two degenerate states. Figure E.4
obtained in this way displays the same characteristic and
non-universal features as
Fig.~E.3. The way in which these features are similar
in both plots corroborates our approach. 
 
\begin{figure}[!t]
\includegraphics[draft=false,keepaspectratio=true,clip,angle=0,%
                   width=
                   \columnwidth]{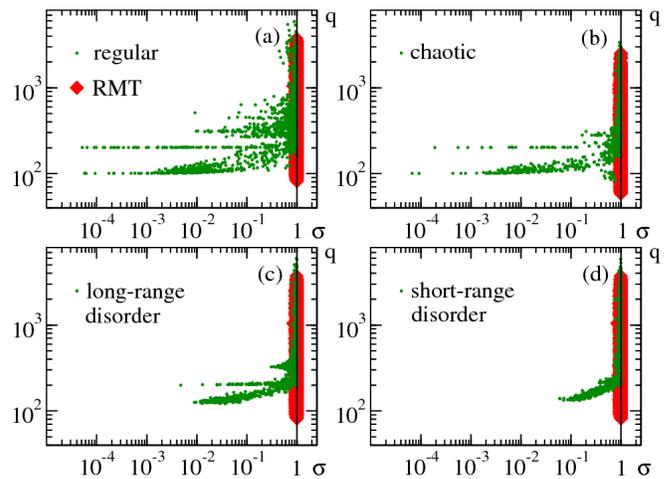}
                   \caption{Same as Fig.~E.3 but with the time-delay eigenvalues $q_i$ and 
                   the noiseless norm $\sigma$ determined from the eigenproblem for the
                   whole Wigner-Smith matrix $Q$ (see text for details). As predicted by our
                   model, a clear correspondence with Fig.~E.3 can be observed for small
                   values of $\sigma$. This indicates that NOTEs as identified by the 
                   procedure proposed in the main text naturally emerge as eigenstates of the
                   Wigner-Smith time-delay matrix. }
\label{fig:4}
\end{figure}

The small differences between Figs.~E.3, E.4 which remain for 
small values of $\sigma$
can be well described with the help of degenerate
perturbation theory (not shown). In this description the
``perturbation'' is given in terms of the off-diagonal matrices
$Q_{12}$, $Q_{21}$ which mix the eigenstates of the ``unperturbed''
matrices $Q_{11}$, $Q_{22}$.  Eigenstates and eigenvalues of $Q_{11}$
thus get perturbed through admixtures from other states which, in
further consequence, lead to the above differences. In turn,
these admixtures should vanish when the perturbation of a
$Q_{11}$-eigenstate $\vec{q}_{i,L}^{\,\,{\rm in}}$ by the matrices
$Q_{12}$ or, equivalently, by $Q_{21}$ goes to zero. Note that this
situation is exactly realized when the null-space norm
$\chi=\|Q_{21}\vec{q}_{i,L}^{\,\,{\rm in}}\|$ (or, equivalently, the
noiseless norm $\sigma$, see discussion above) goes to zero. This
reasoning not only substantiates the fact that the norms $\chi$ and
$\sigma$ are a suitable means to measure the quality of a NOTE but
also that in the numerically unreachable limit of infinitely many lead
modes ($N\to\infty$)
we should expect the Wigner-Smith time-delay matrix to feature exactly
degenerate pairs of NOTEs.

\section{C.~NOTEs and classical trajectory bundles}

An interesting question to ask is whether all NOTEs we find are
associated with classical trajectory bundles and vice versa. To answer
this question we have performed a number of tests: our work builds on
the established result from the literature (see, e.g.,
\cite{schomerus}) according to which phase space bands with a size
larger than Planck's constant give rise to noiseless scattering
channels. In the weakly disordered regime (which we consider) each of
these bands can be associated with a bundle of classical trajectories
of equivalent topology and similar length (a situation which is in
general not fulfilled for strongly disordered systems).
To verify whether all NOTEs are associated with such classical
trajectory bundles we thus need to test if the NOTEs generated with
our procedure [see conditions (i),(ii) on page 3 in the
main paper text] are composed exclusively of noiseless scattering
channels. As demonstrated in section B (see, in particular
Figs.~E.1, E.2) all NOTEs which we find in the ballistic structures
(regular and chaotic), indeed, have this property. We also find that
in the weakly disordered structures where such trajectory bundles are
disturbed, NOTEs are of considerably lower quality. Furthermore, the
quality and quantity of NOTEs correlates strongly with the area of a
given phase space band: consider, e.g., that no NOTEs are found with
very large time-delays (see Figs.~E.3, E.4) as these are associated with
very small phase space bands. We conclude from these observations that
all NOTEs we find can, indeed, be associated with classical trajectory
bundles. 
 
The above arguments still leave open whether each NOTE can be 
associated with only a single or multiple 
trajectory bundles. Since NOTEs are time-delay eigenstates and 
different bundles have, in general, different time-delays our 
procedure should yield only a single bundle per NOTE already 
by construction. To test whether this expectation is, indeed, 
fulfilled we performed the following additional tests:
We have checked that each time-delay eigenvalue $q_i$ associated with 
a NOTE in the regular cavity corresponds to a classical 
time-delay value from a single classical phase space band:
Based on the classical dynamics in a rectangular cavity the product of
a proper delay time $q_i$ with the eigenstate's average longitudinal
velocity, $\langle|v^i_x|\rangle=\hbar \langle |k^i_x|\rangle/m$, must
be an integer multiple of the cavity width $W$, corresponding to the
number $M$ of the state's left-right cavity traversals,
$q_i\,|v^i_x|=M W$. With the average longitudinal momentum
$\langle|k^i_x|\rangle$ of state $i$ being determined as follows,
$\langle|k^i_x|\rangle=\sum_{n=1}^N|(\vec{q}_{i,L}^{\,\,{\rm
    in}})^{\phantom{i}}_n|^2 \sqrt{k^2-(n\pi/d)^2}$, we find this
criterion to be fulfilled with a relative error below $2\%$ by all the
NOTEs for which $\chi\lesssim 20$. This result confirms that for the
rectangular cavity our numerical procedure yields proper delay times
in very good agreement with classical expectation. In the chaotic
cavity we have checked that NOTEs which have no overlap with the
circular boundary feature a one-to-one correspondence with the
equivalent states in the regular cavity (both in terms of the
time-delay eigenvalue and in terms of the wave function).  For the
remaining NOTEs in the chaotic cavity (which do have an overlap with
the circular boundary) and for all NOTEs in the disordered cavities we
have checked explicitly with the help of the corresponding scattering
wave functions that these NOTEs all feature highly collimated wave
function patterns as expected for states that are supported by
classical trajectory bundles. Details can be found in section E, where
we have provided the wave function patterns of NOTEs on specific phase
space bands (in the ballistic cavities) and all NOTEs with
$\chi\lesssim 30$ in the disordered cavities.

Whereas the above tests suggest that each NOTE we found is
associated with a single phase space band/trajectory bundle this
association does not always hold in the opposite direction: as stated
above, NOTEs can only be formed on phase space bands with a size
larger than $h$ (a condition which we have explicitly verified in an
earlier paper \cite{rotter2}). A further restriction to form NOTEs on
such noiseless channels comes from the well-defined time-delay
associated with NOTEs.  This requirement imposes an additional
constraint which makes it harder for NOTEs to fit on a given phase
space band as compared to the transmission eigenstates of $t^\dagger
t$.  Correspondingly, we find that the transmission of NOTEs is
generally not as noiseless as that of the transmission eigenstates
(not shown). 

We mention, parenthetically, that imperfections both in the formation
of noiseless channels and NOTEs
are mostly reflected by those states labeled as ``intermediate'' in
Fig.~E.1a: these states deviate from RMT statistics but do not have a
good quality as NOTEs. Such intermediate states are found to persist
in the presence of weak disorder (Fig.~E.1c,d).

\section{D.~WKB/eikonal ansatz for NOTEs}
In the main part of the paper we showed that NOTEs are defined by the
two conditions (i) $Q_{11}\,\vec{q}_{i,L}^{\,\,{\rm in}}=
q_i\,\vec{q}_{i,L}^{\,\,{\rm in}}$ and (ii) $Q_{21}\,
\vec{q}_{i,L}^{\,\,{\rm in}}=\vec{0}$. In this part of the appendix we
investigate how these equations can be satisfied, based on the
observation from section C that NOTEs are time-delay eigenstates
located on individual phase space bands. We assume, for simplicity,
that we are dealing with a transmission band, leading to a fully
transmitted NOTE. Our arguments can, however, be applied to reflection
bands as well. 

Consider first the eigenproblem in (i): 
$Q_{11}\,\vec{q}_{i,L}^{\,\,{\rm in}}=i\hbar\left(\dot{r}^\dagger
  r+\dot{t}^\dagger t\right) \,\vec{q}_{i,L}^{\,\,{\rm
    in}}=q^{\phantom{i}}_i\,\vec{q}_{i,L}^{\,\,{\rm in}}$.  Using the
property that $\vec{q}_{i,L}^{\,\,{\rm in}}$ is a fully transmitted
noiseless state, $t^\dagger t \,\vec{q}_{i,L}^{\,\,{\rm
    in}}=\vec{q}_{i,L}^{\,\,{\rm in}}$, we have
$\dot{r}^\dagger r \,\vec{q}_{i,L}^{\,\,{\rm
    in}}=\dot{r}^\dagger\,\vec{0}=\vec{0}$, leaving us with
$Q_{11}\,\vec{q}_{i,L}^{\,\,{\rm in}}=i\hbar\,\dot{t}^\dagger
t\,\vec{q}_{i,L}^{\,\,{\rm in}}=q_i\,\vec{q}_{i,L}^{\,\,{\rm in}}$ for
the evaluation of which we perform a singular value decomposition
(SVD) of $t=\sum_i t_i$ with
$t_i=\vec{u}_i\sigma_i\vec{v}^\dagger_i$. Here the right-singular
vectors $\vec{v}_i=\vec{\tau}_{i}^{\,\,{\rm in}}$ are the eigenvectors
of $t^\dagger t$, the left-singular vectors
$\vec{u}_i=\left(\vec{\tau}_{i}\,\!'^{\,{\rm in}}\right)^*$ are the
eigenvectors of $tt^\dagger=\left(t'^\dagger t'\right)^*$, and the
singular values are the square roots of the corresponding transmission
eigenvalues, $\sigma_i=\sqrt{\tau_i}\in[0,1]$. In wave transport
through complex scattering landscapes the non-deterministic singular
values $0<\sigma_i<1$ and their corresponding singular vectors
$\vec{v}_i$ and $\vec{u}_i$ feature a very strong energy dependence as
reflected, e.g., in the familiar conductance fluctuations. In the
degenerate noiseless subspace where $\sigma_i=0,1$ the situation is
more complicated: states which are noiseless due to an appropriate
interference of path contributions (like in Fig.~\ref{fig:1}a) also
feature a significant energy dependence induced by multi-path
interference. For those states that are noiseless due to a resolution
of the underlying classical phase space (like in Fig.~\ref{fig:1}b)
the singular values $\sigma_i$ all stay at the energy-independent
degenerate values of $\sigma_i=0,1$ (see, e.g.,
\cite{schomerus}). Since the corresponding singular vectors are,
typically, distributed over several transmission (reflection) bands
for $\sigma_i=1$ ($\sigma_i=0$) their energy-dependence is not well
defined unless we fix this distribution. As we argue above, a
distribution where each state is located on a single band can be
achieved by demanding that the noiseless vectors $\vec{v}_i$ are
NOTEs, $\vec{v}_i=\vec{q}_{i,L}^{\,\,{\rm in}}$ and
$\vec{u}_i=t\,\vec{q}_{i,L}^{\,\,{\rm in}}$. On a single band all
involved trajectories will accumulate a very similar phase, leading to
energy-independent constructive interference such that the
incoming/right singular vectors have only a very weak energy
dependence, $\partial_E \vec{v}_i\approx\vec{0}$. In this relation the
arbitrary global phase that we are free to choose in any SVD was fixed
for the vector $\vec{v}_i$ such as to minimize its
energy-dependence. This condition, in turn, also determines the
corresponding outgoing/left singular vector,
$\vec{u}_i=\vec{u}^{\,0}_i\,e^{i\phi_b(E)}$, to consist of an
energy-independent part, $\partial_E \vec{u}^{\,0}_i\approx\vec{0}$,
and an energy-dependent phase shift, $e^{i\phi_b(E)}$. The phase shift
$\phi_b(E)={\mathcal S}_b(E)/\hbar-\pi\mu_b/2$ can be associated with
the action ${\mathcal S}_b(E)=\int_b \vec{k}\,d\vec{l}$ and the Maslov
index $\mu_b$ accumulated along the phase space band $b$.
With the transmission amplitudes for the transmitted NOTEs now being
$t_i=\vec{q}_{i,R}^{\,\,{\rm in}\,*}e^{i\phi_b}\vec{q}_{i,L}^{\,\,{\rm
    in}\,\dagger}$, we can further simplify condition (i):
$Q_{11}\,\vec{q}_{i,L}^{\,\,{\rm in}}=i\hbar\,\dot{t}^\dagger
t\,\vec{q}_{i,L}^{\,\,{\rm in}}=i\hbar\left[\partial_E(t^\dagger
  t\,\vec{q}_{i,L}^{\,\,{\rm in}})-t^\dagger\partial_E
  (t\,\vec{q}_{i,L}^{\,\,{\rm in}})\right]=(\partial_E {\mathcal
  S}_b)\, \vec{q}_{i,L}^{\,\,{\rm in}}$, from which we obtain the
time-delay eigenvalue $q_i=\partial_E {\mathcal S}_b$, in full analogy
to the one-dimensional case \cite{Reichl2004Transition}. Similarly,
condition (ii) reads: $Q_{21}\,\vec{q}_{i,L}^{\,\,{\rm
    in}}=i\hbar\,\dot{r}'^\dagger t\,\vec{q}_{i,L}^{\,\,{\rm
    in}}=i\hbar\left[\partial_E(r'^\dagger t\,\vec{q}_{i,L}^{\,\,{\rm
      in}})-r'^\dagger\partial_E (t\,\vec{q}_{i,L}^{\,\,{\rm
      in}})\right]=i\hbar\left[\partial_E(r'^\dagger
  t\,\vec{q}_{i,L}^{\,\,{\rm in}})-(\partial_E{\mathcal
    S}_b)\,r'^\dagger t\,\vec{q}_{i,L}^{\,\,{\rm in}}\right]$, which
expression must be zero since the unitarity of the $S$-matrix implies
$r'^\dagger t =-t'^\dagger r$ and $t'^\dagger
r\,\vec{q}_{i,L}^{\,\,{\rm in}}=0$ for a noiseless transmission state.

We have thus verified that WKB/eikonal-type scattering states which 
are characterized by their phase shift accumulated along a given
trajectory bundle satisfy the two defining equations (i),(ii)
which we have put forward to construct NOTEs. 

\section{E.~Wave function images}
In the figures attached below we provide a number of wave function
images as obtained for the four different cavity types discussed in
the main article. In the regular and chaotic cavity we show all states
belonging to a specific phase space band: Figs.~E.5, E.6 illustrate
how NOTEs systematically increase their transverse quantization until,
when a band is filled, overlap with phase space regions outside this
band lead to a significant drop in the quality of NOTEs.  For the
disordered cavities we display all wave function patterns with a
quality below $\chi\lesssim 30$: Figs.~E.7, E.8 show that all states of
this quality still feature very collimated wave functions in spite of
the presence of weak disorder. Note that in the cavity with long-range
correlations in the disorder the number of NOTEs of this quality is
much higher than with short-range disorder.

\begin{figure*}[!t]
\includegraphics[draft=false,keepaspectratio=true,clip,%
                   width=
                   0.95\textwidth]{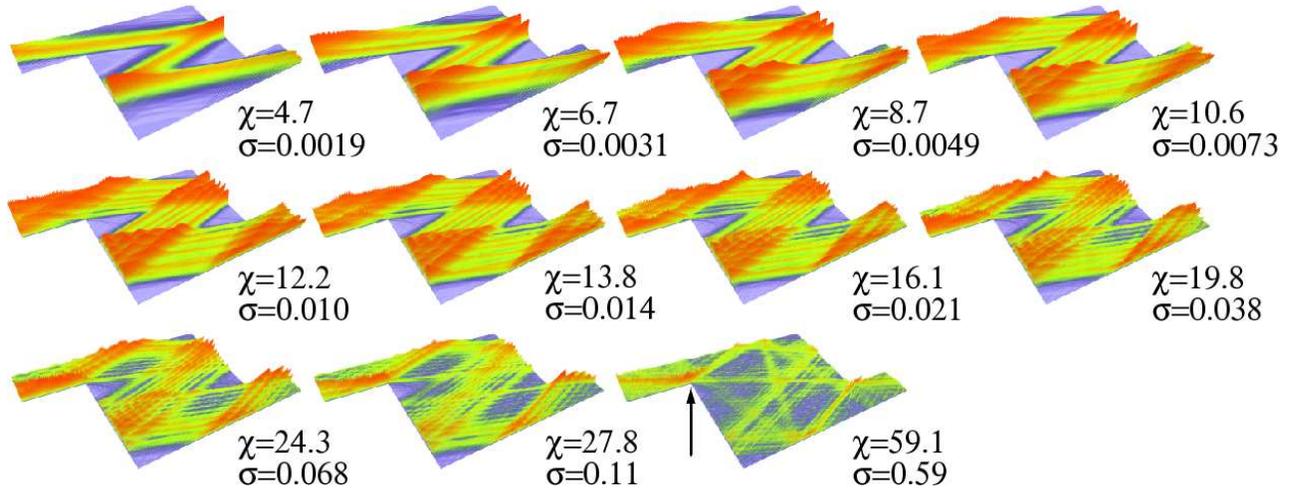}
                   \caption{Wave function densities of NOTEs in the
                     rectangular cavity at wave number $k=75.5\pi/d$:
                     States associated with the transmission band T2
                     (see Fig.~1c in the main text) are shown with
                     their respective null-space and noiseless norms
                     $\chi,\sigma$. With increasing transverse
                     quantization the quality of states (as measured
                     by $\chi,\sigma$) deteriorates
                     monotonically. This decrease in the quality of
                     NOTEs is accompanied by the appearance of
                     diffraction effects at the sharp lead mouth (see
                     arrow in the bottom row).}
\label{fig:5}
\end{figure*}

\begin{figure*}[!t]
\includegraphics[draft=false,keepaspectratio=true,clip,%
                   width=
                   0.95\textwidth]{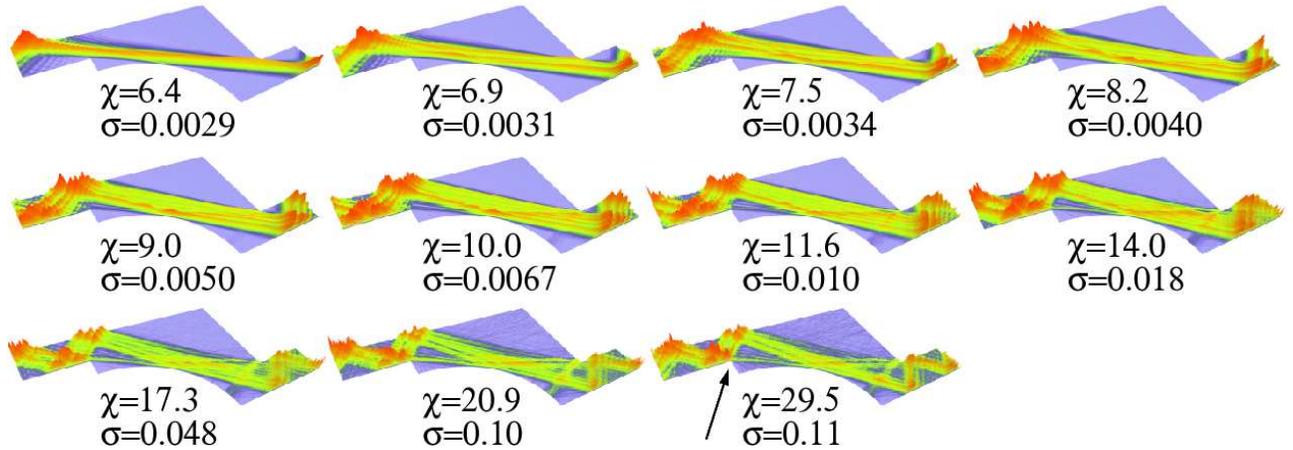}
                   \caption{Wave function densities of NOTEs in the
                     chaotic cavity at wave number $k=75.5\pi/d$:
                     States associated with the direct transmission
                     path are shown with their respective null-space
                     and noiseless norms $\chi,\sigma$. With
                     increasing transverse quantization states have
                     more overlap with the circular boundary part. For
                     low-quality NOTEs diffraction effects are
                     observed also here (see arrow in the bottom
                     row). }
\label{fig:7}
\end{figure*}

\begin{figure*}[!t]
\includegraphics[draft=false,keepaspectratio=true,clip,%
                   width=
                   0.95\textwidth]{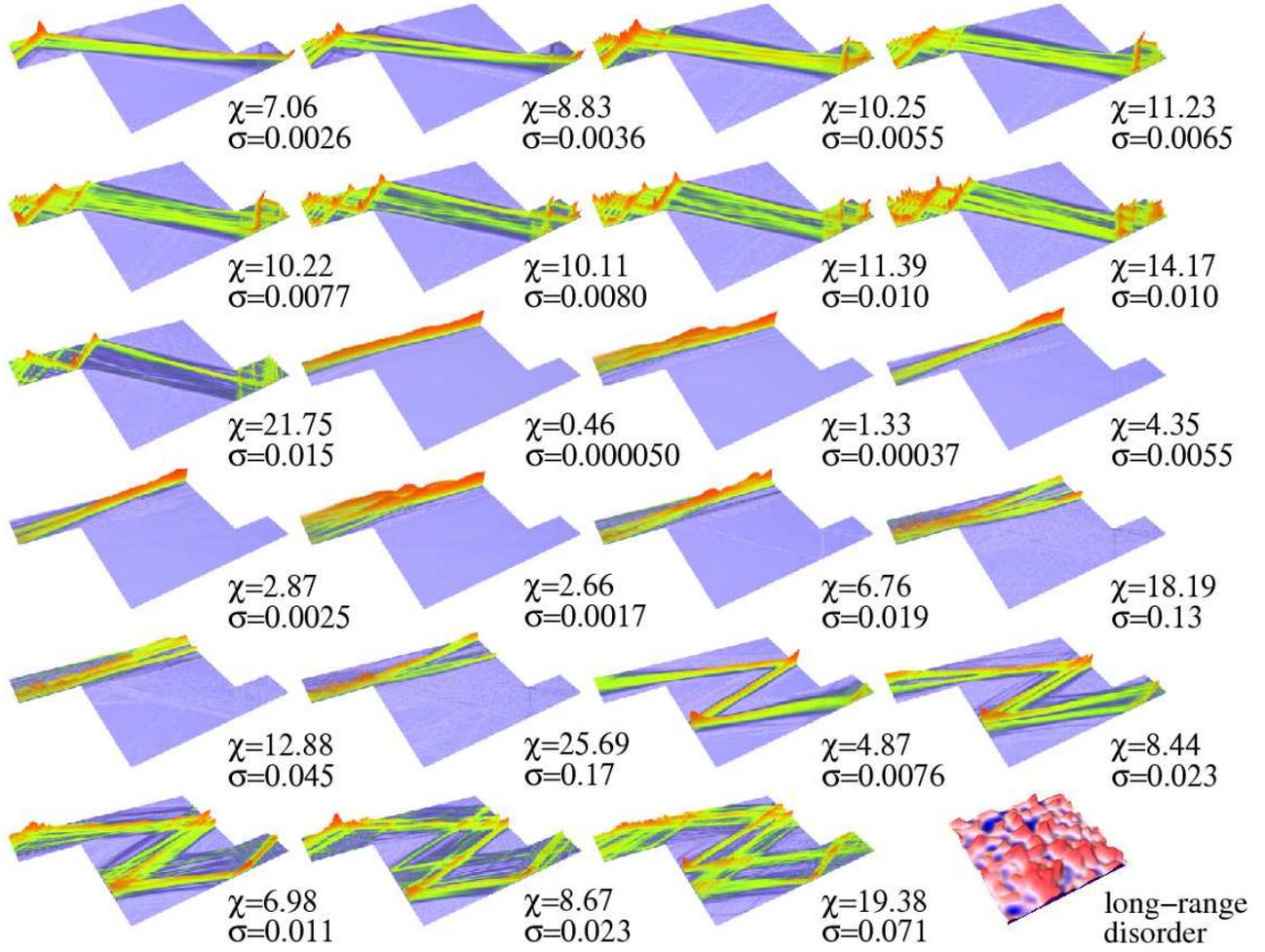}
                   \caption{Wave function densities of NOTEs in the
                     disordered cavity with long-range disorder at
                     wave number $k=75.5\pi/d$: all NOTEs with a
                     null-space norm $\chi\lesssim 30$ are shown in
                     the order of their time-delay value $q_i$. The
                     number of states with this quality is here much
                     larger than in the case of short-range disorder
                     with the same disorder amplitude, $V_0=10\% E$
                     (see Fig.~E.8 for comparison). The employed
                     potential landscape with disorder correlation
                     length $kr_c=30\pi$ is shown in the rightmost
                     panel in the bottom row.  }
\label{fig:8}
\end{figure*}

\begin{figure*}[!t]
\includegraphics[draft=false,keepaspectratio=true,clip,%
                   width=
                   0.95\textwidth]{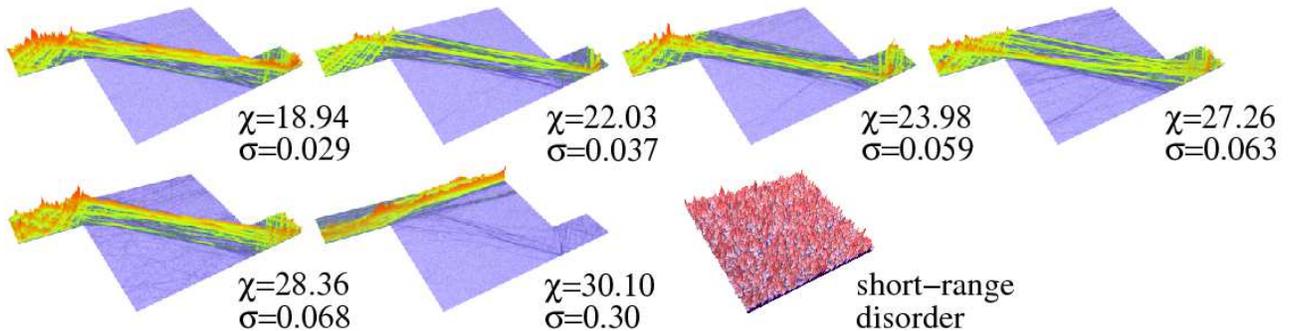}
                   \caption{Wave function densities of NOTEs in the
                     disordered cavity with short-range disorder at
                     wave number $k=75.5\pi/d$: all NOTEs with a
                     null-space norm $\chi\lesssim 30$ are shown in
                     the order of their time-delay value $q_i$. The
                     disorder amplitude ($V_0=10\% E$) is here the
                     same as in Fig.~E.7, but the number of states
                     with quality $\chi\lesssim 30$ is strongly
                     reduced as compared to the case of long-range
                     disorder. The employed potential landscape with
                     disorder correlation length $kr_c=5\pi$ is shown
                     in the rightmost panel in the bottom row.  }
\label{fig:9}
\end{figure*}